\documentclass[10pt]{article}
\usepackage{graphicx, amsmath, amssymb, amsthm, hyperref}
\usepackage[a4paper, margin=1in]{geometry}
\usepackage{subcaption}

\title{Resolution Estimation of a Digital Holographic Microscope Using Neural Network Analysis of Reconstructed Images}
\author{Arthur G. Fedorov\\ \small North-Eastern Federal University, 58 Belinsky Str., Yakutsk, 677000 Russia\\ \small Corresponding author: \texttt{ag.fedorov@s-vfu.ru}}
\date{}

\begin{document}
\maketitle

\begin{abstract}
This paper presents a method for estimating the resolution of a digital holographic microscope based on the analysis of reconstructed images using neural networks. The spectral bandwidth of the source ($\Delta \lambda$) is used as a controlled degradation parameter.

Numerical simulations were performed within inline Gabor holography. A dataset of reconstructed images was generated for several test objects over a $\Delta \lambda$ range from 0.05 to 20 nm.

The model predicts $\Delta \lambda$ from the reconstructed images with high precision. The predictions are consistent with standard resolution metrics, including FWHM, MTF, and the USAF resolution criterion.

The generalization analysis shows that the model is sensitive to the type of degradation. It correctly captures interferometric distortions and responds selectively to the underlying physical mechanism.

The proposed approach allows resolution estimation without explicit modeling of all degradation factors and can be used for compact holographic systems.
\end{abstract}


\section{Introduction}

Inline digital holography [1–3] is one of the simplest and most versatile configurations for recording and reconstructing object structure. In its basic form, the system consists of a radiation source, an object, and a detector [4], while image reconstruction is performed numerically [5–7]. Unlike conventional optical microscopes based on complex and expensive lens systems, digital holography enables the development of compact, portable [8–11], and low-cost devices [12] with integrated software for image reconstruction and analysis [13–15].

The simplicity of the inline configuration (source–object–detector) makes it attractive for both fundamental research and engineering applications. Such systems are widely used in electron microscopy [16–18], where holography is applied to study structural features of objects, as well as in optical digital holography for biomedical diagnostics [19,20], microfluidics [21], and field measurements [8,9].

Recent trends in scientific instrument miniaturization have stimulated the development of compact and portable digital holographic microscopes [8,12,22,23]. Such systems can be built using standard components and provide a significant cost reduction compared to commercial solutions [12]. In particular, implementations based on low-cost laser diodes, LED sources, CMOS sensors, and simple printed circuit boards have been reported. The cost of a complete laboratory prototype in such systems often does not exceed 150 USD [23].

However, the low cost of the components is also one of the main limitations of these systems. The quality of the coherent source, detector characteristics, system geometry, and limited field of view all affect the final resolution of a digital holographic microscope. Separating the influence of these factors analytically is difficult. Although each factor has its own spectral signature, their effects are combined in the reconstructed image, and the resolution is ultimately limited by the factor with the lowest cutoff frequency. As a result, the contribution of each individual factor becomes inseparable from the overall image degradation. This makes the analysis of individual factors inefficient and requires approaches capable of working with their combined effect.

In this work, a method is proposed for estimating system resolution without explicitly considering a specific limiting factor. The approach is based on deep learning methods that operate on the combined effect of image degradation without explicit modeling of individual system parameters.

As a demonstration case, the influence of source temporal coherence on the resolution of a digital holographic microscope is considered, since this factor produces a monotonic and well-controlled degradation of the reconstructed image quality. The applicability limits of the method and the transferability of the model to different object types are also analyzed.

\section{Methods}
\subsection{Holographic configuration and object parameters}
Hologram simulation and image reconstruction were performed within inline Gabor holography (Fig. 1). The input field was assumed to be a plane wave $U_{\mathrm{in}}$. After interaction with the object, the object wave $U_{\mathrm{obj}}$ is formed, while part of the field propagates without changes and forms the reference wave $U_{\mathrm{ref}}=U_{\mathrm{in}}$. The interference between $U_{\mathrm{obj}}$ and $U_{\mathrm{ref}}$ in the detector plane produces the holographic image $U_{\mathrm{scr}}$. Wave propagation was modeled numerically in this work [24].
\begin{figure}[h]
\centering 
\includegraphics[width=0.5\linewidth]{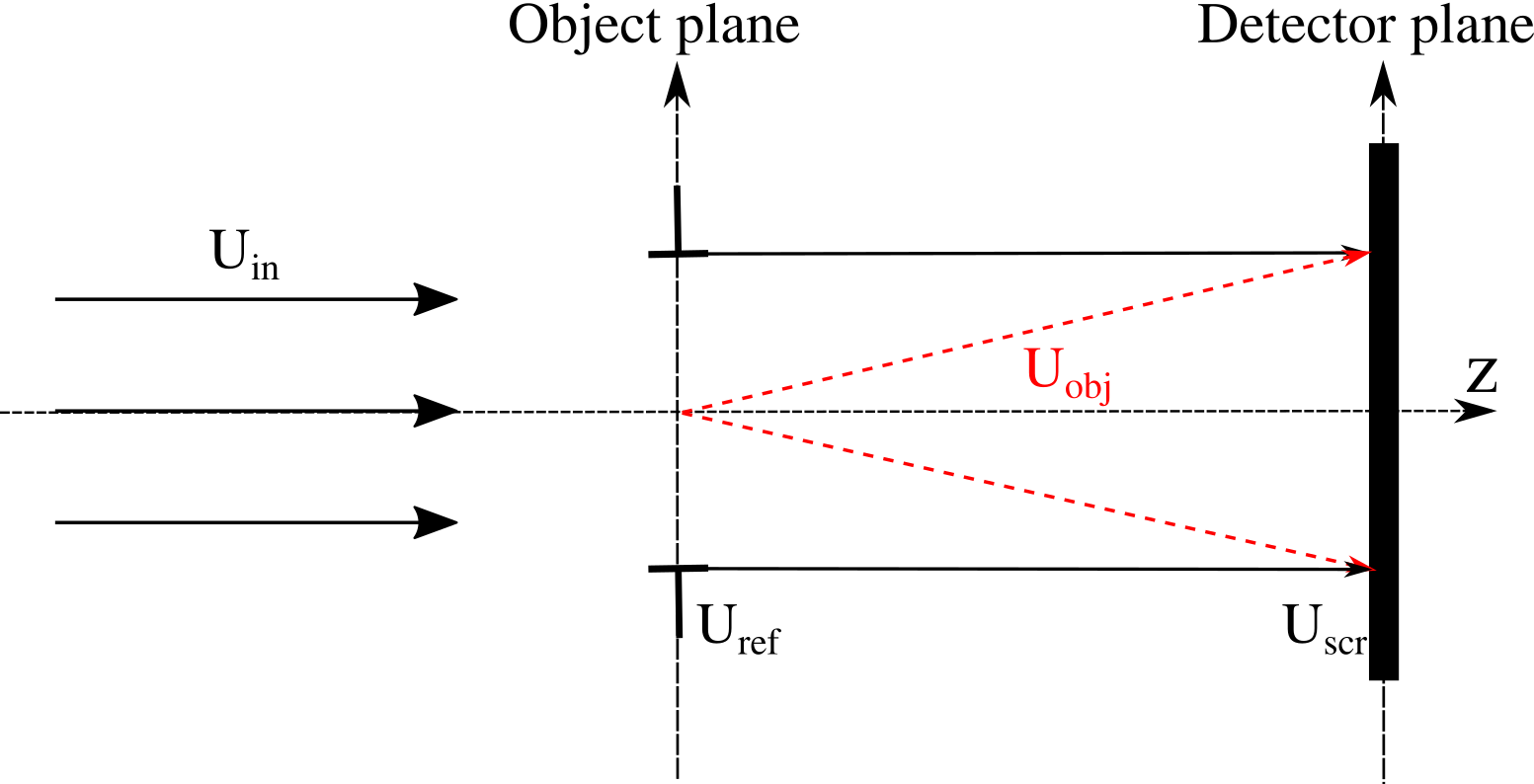} 
\caption{General scheme of the holographic configuration.}
\end{figure}

The dataset included four types of test objects: a chirp pattern, a sinusoidal grating, a star pattern, and a USAF resolution target (Fig. 2). All objects were binary amplitude images with different spatial-frequency characteristics. The selected set of objects allows the robustness of the method to be analyzed for different types of spatial structure under fixed degradation conditions. Simulations were performed for square images of size N×N, where N=256.
\begin{figure}[h]
\centering
\begin{subfigure}{0.2\linewidth}
    \includegraphics[width=\linewidth]{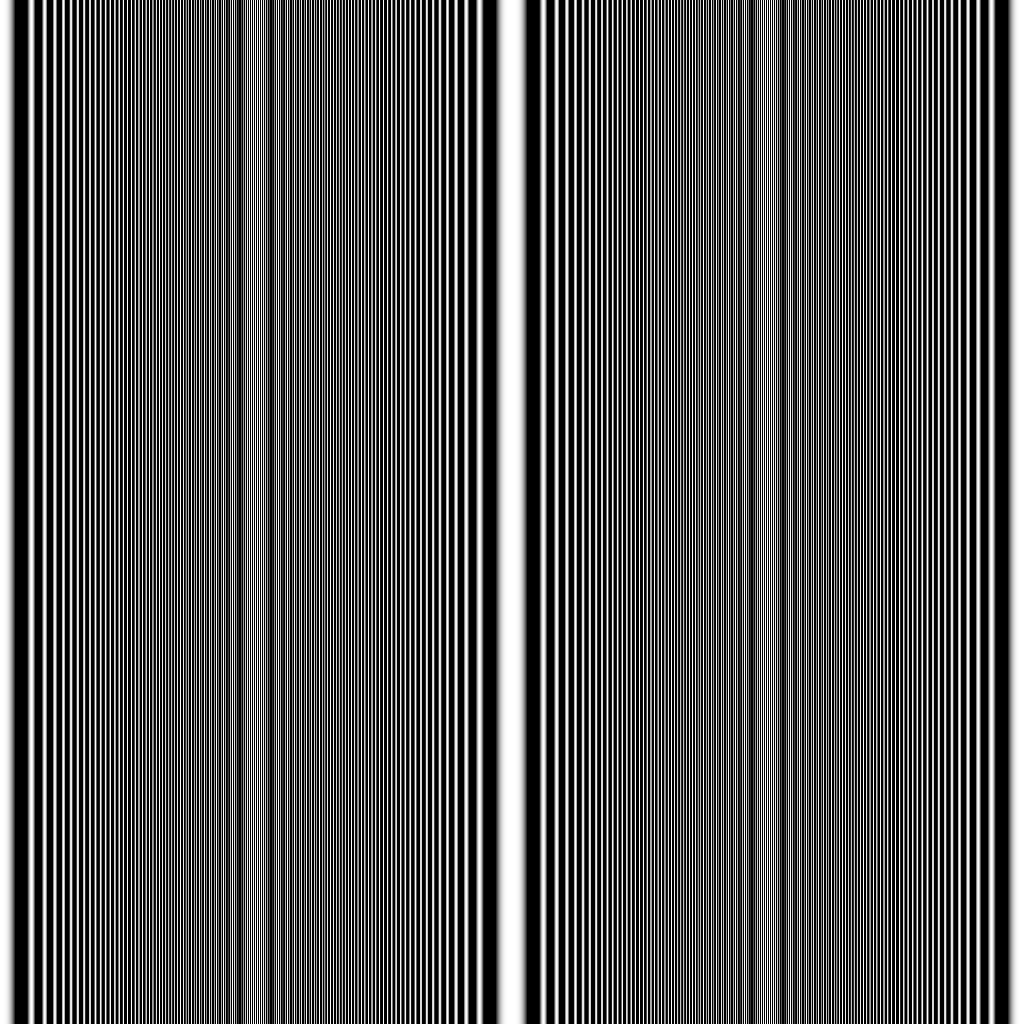}
    \caption{}
\end{subfigure}
\hfill
\begin{subfigure}{0.2\linewidth}
    \includegraphics[width=\linewidth]{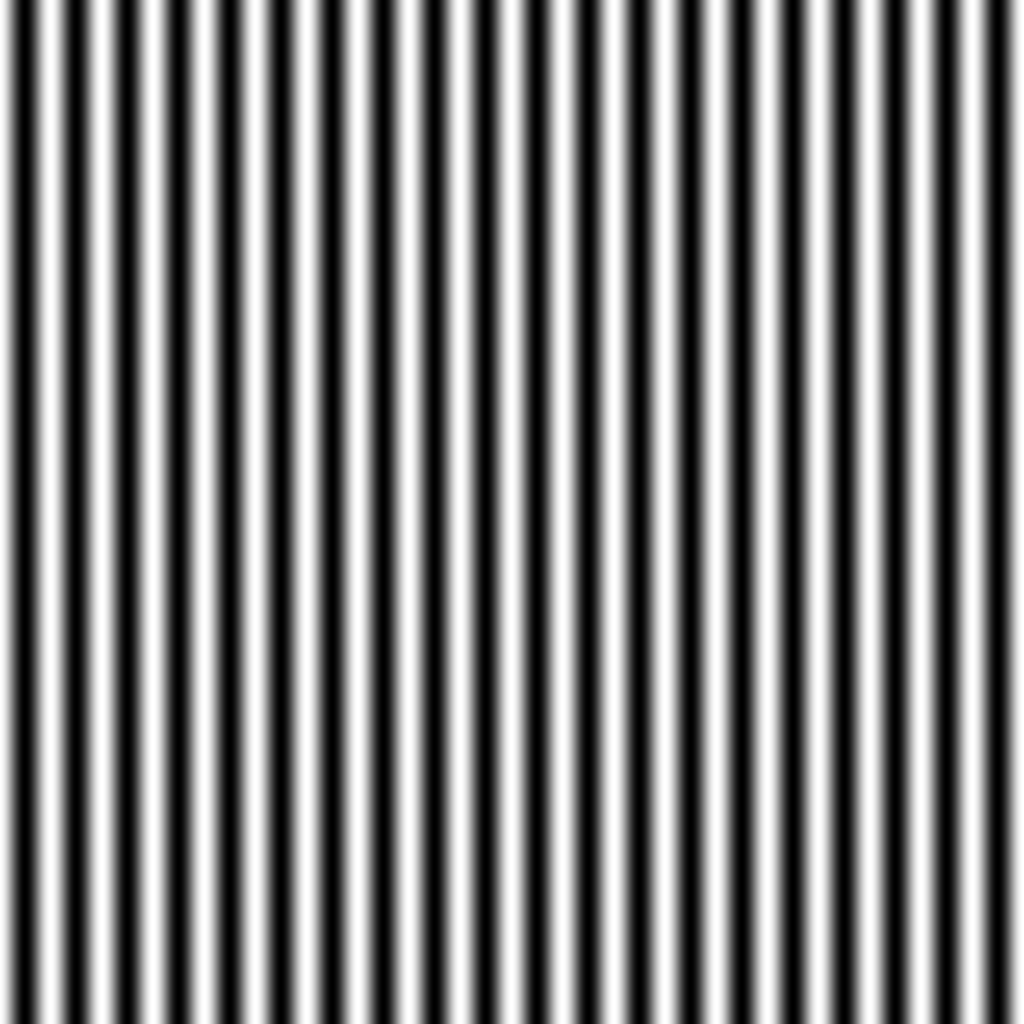}
    \caption{}
\end{subfigure}
\hfill
\begin{subfigure}{0.2\linewidth}
    \includegraphics[width=\linewidth]{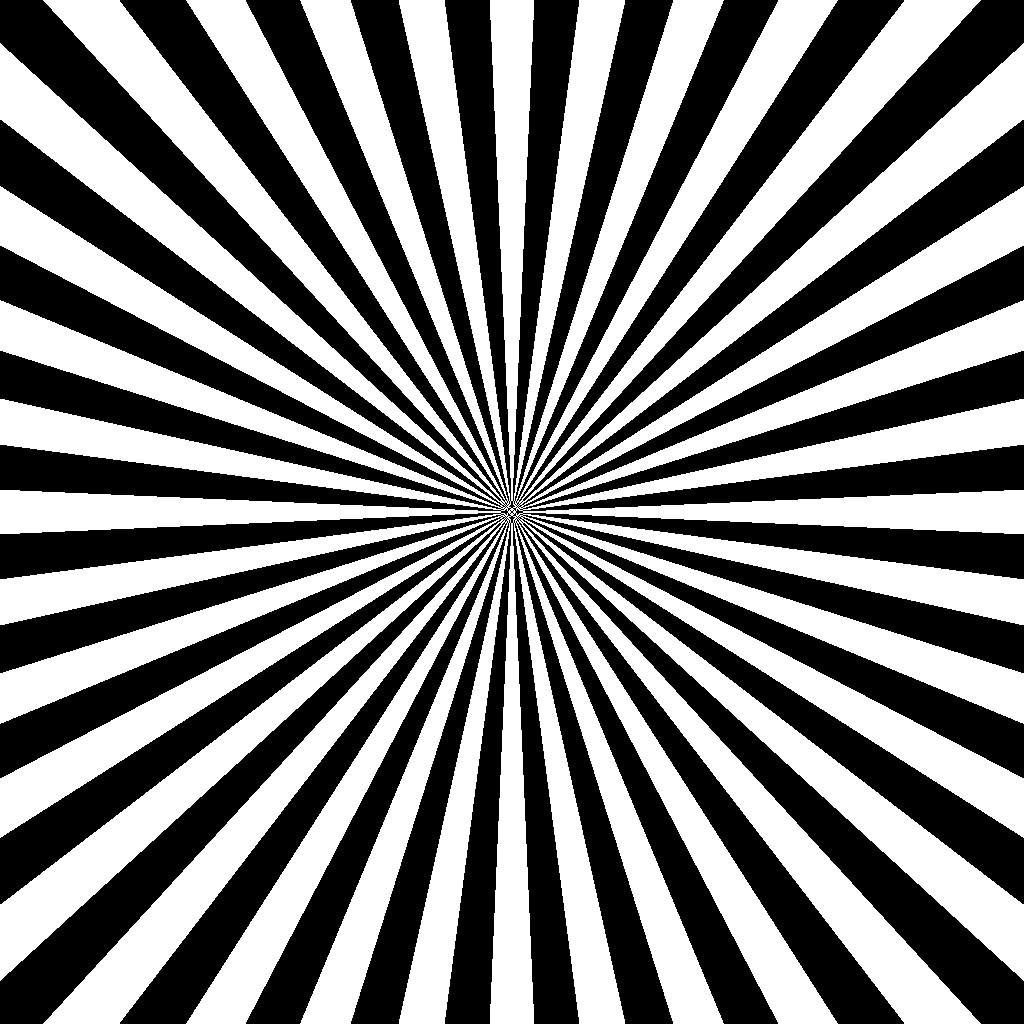}
    \caption{}
\end{subfigure}
\hfill
\begin{subfigure}{0.2\linewidth}
    \includegraphics[width=\linewidth]{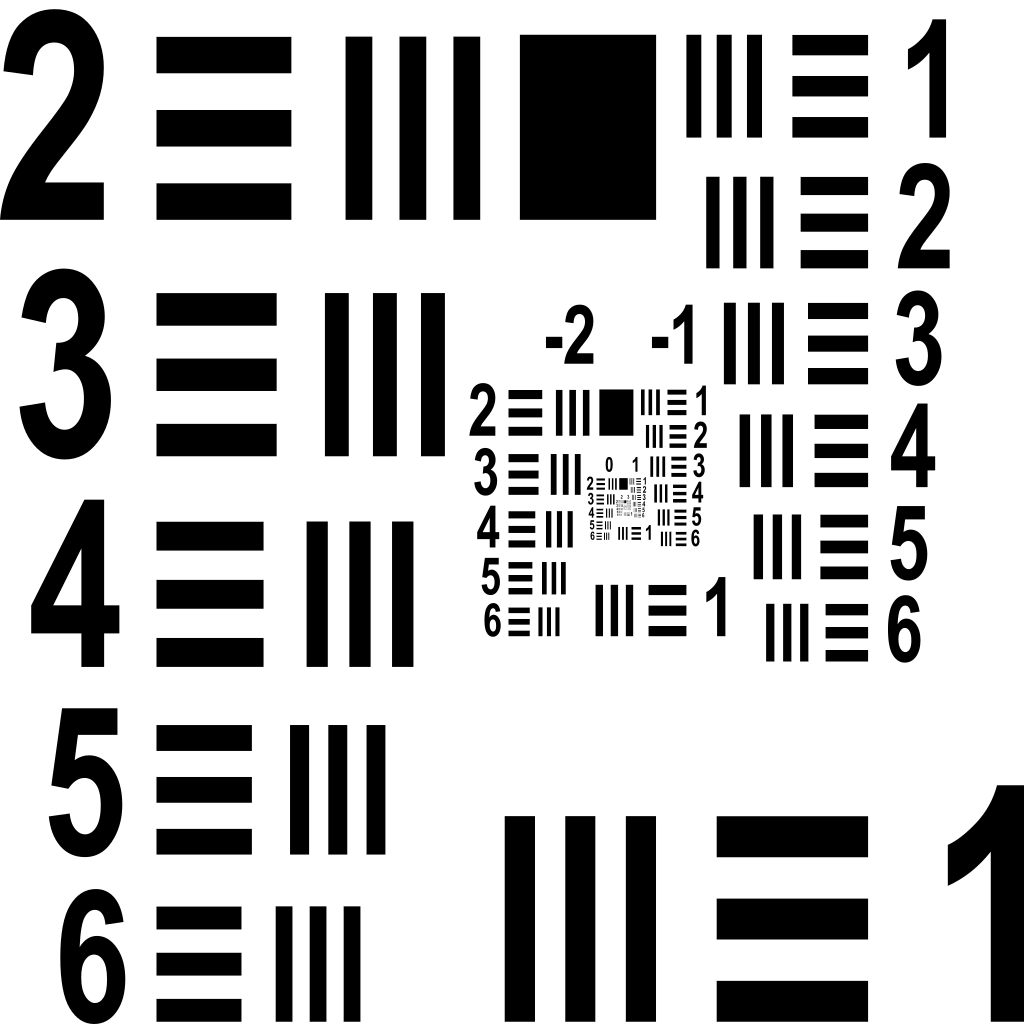}
    \caption{}
\end{subfigure}
\caption{Test objects used in the simulations: (a) chirp pattern, (b) sinusoidal grating, (c) star pattern, and (d) USAF resolution target.}
\end{figure}

\subsection{Hologram simulation and reconstruction}

Hologram simulation was based on the numerical solution of the Rayleigh–Sommerfeld integral [25,24,6]:
\[
U_{\mathrm{scr}} = U_{\mathrm{ex}} \otimes h,
\]
where $U_{\mathrm{ex}}=U_{\mathrm{ref}}+U_{\mathrm{obj}}$, h is the impulse response of the system, and $\otimes$ denotes the convolution operation.

Numerical simulations were performed using the angular spectrum method:
\[
U_{\mathrm{scr}}(x_0,y_0) \approx \mathcal{F}^{-1} \left\{\mathcal{F} \left[U_{\mathrm{ex}}(x,y)\right]H(f_X,f_Y)\right\},
\]
where $H(f_X,f_Y)$ is the transfer function, and $\mathcal{F}$ and $\mathcal{F}^{-1}$ are the forward and inverse Fourier transforms.

Numerical reconstruction was performed using the back-propagation method:
\[
I(x_0,y_0) \approx  \left| \mathcal{F}^{-1} \left\{ \mathcal{F} \left[ U_{\mathrm{scr}}(x_0,y_0) \right] H(f_X,f_Y) \right\} \right|,
\]
where $I(x_{0},y_{0})$ is the image of the reconstructed object. Figure 3 shows an example of hologram simulation and reconstruction for one of the considered objects (USAF resolution target).
\begin{figure}[h]
\centering
\begin{subfigure}{0.4\linewidth}
    \centering
    \includegraphics[width=\linewidth]{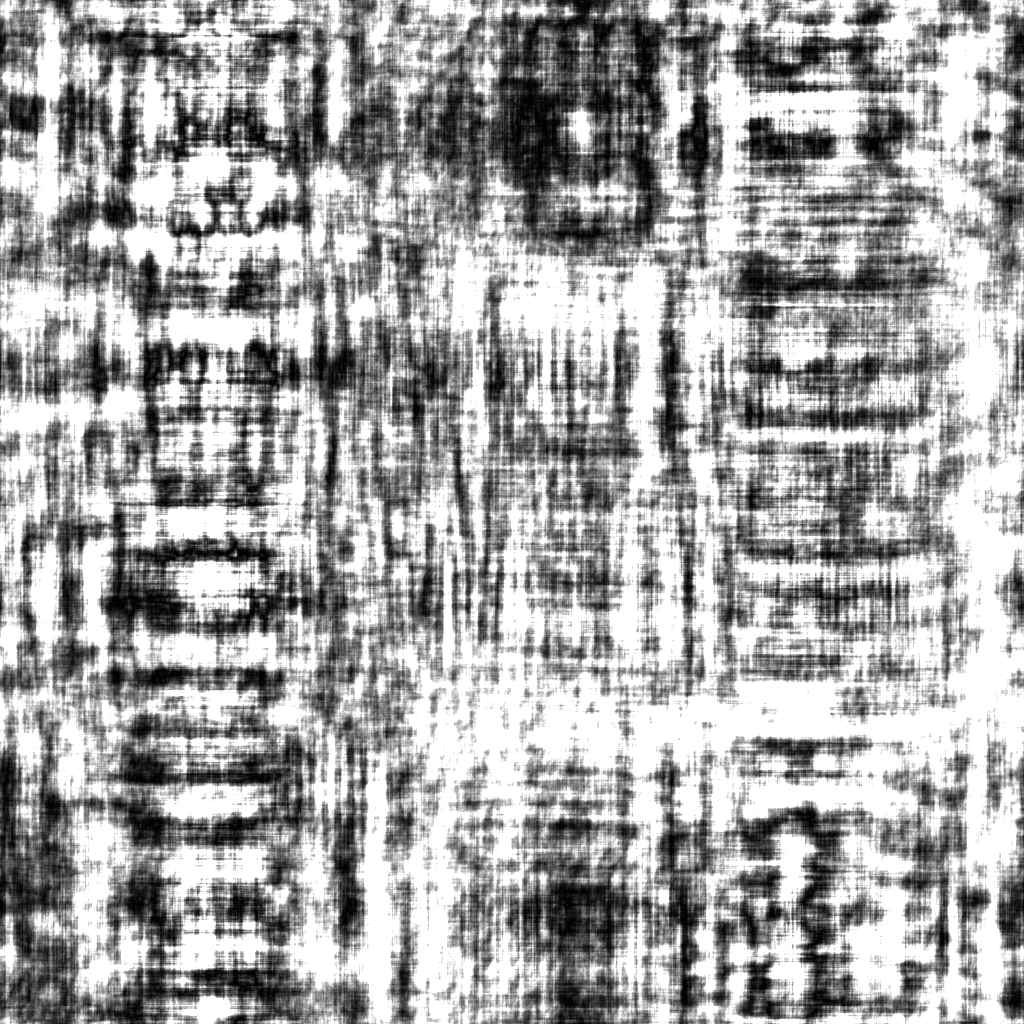}
    \caption{}
\end{subfigure}
\hspace{0.02\linewidth}
\begin{subfigure}{0.4\linewidth}
    \centering
    \includegraphics[width=\linewidth]{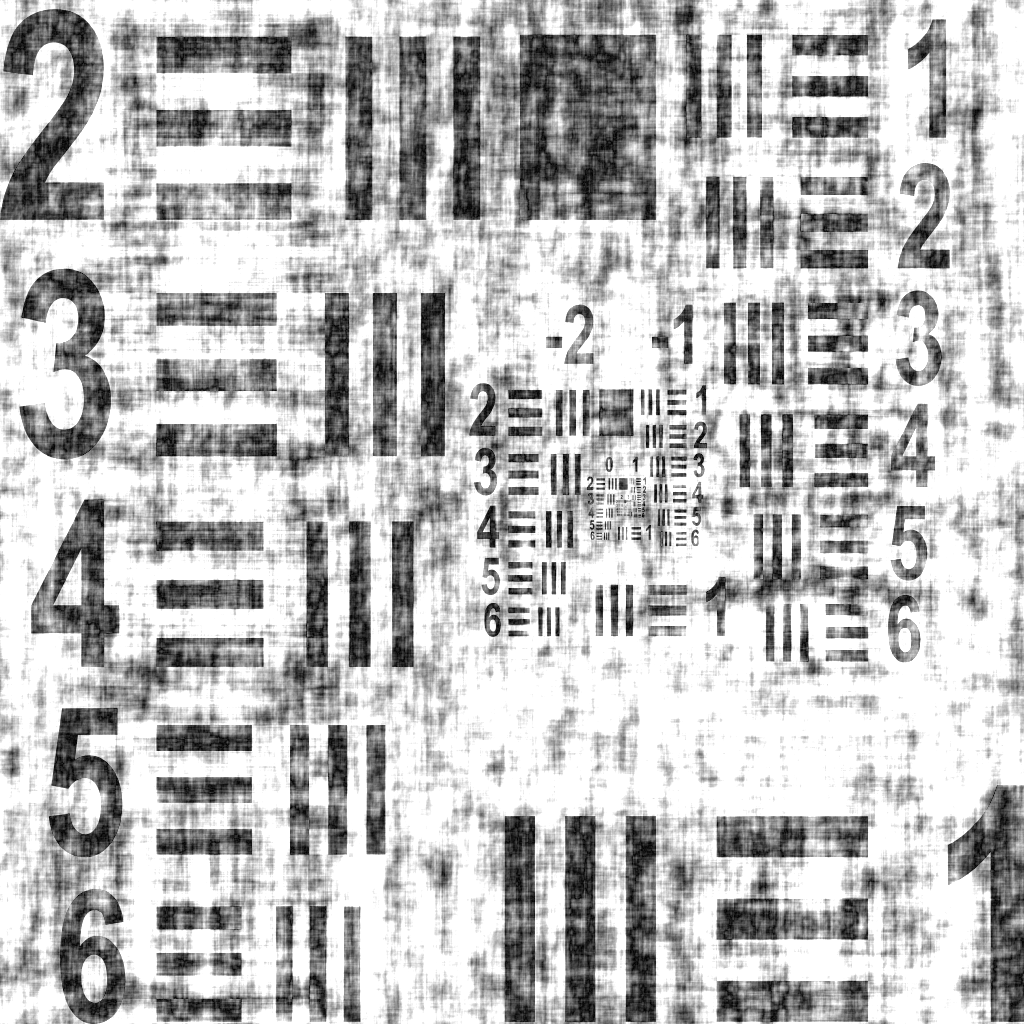}
    \caption{}
\end{subfigure}
\caption{Example of hologram processing: (a) simulated hologram; (b) reconstructed image.}
\end{figure}

\subsection{Dataset generation}
For dataset generation, the spectral bandwidth of the illumination source $\Delta \lambda$ was used as a controlled degradation parameter. This parameter is an intrinsic property of real optical sources. An increase in $\Delta \lambda$ leads to spectral averaging of the interference pattern and, consequently, to a reduction in the contrast of high-frequency spatial components in the reconstructed image.

The effect of finite temporal coherence was modeled during hologram formation. For each value of $\Delta \lambda$, the hologram was calculated as a weighted spectral average of the intensity over a set of wavelengths  distributed around the central wavelength $\lambda_{0}=630$ nm. The source spectrum was assumed to have a Gaussian profile, and the spectral standard deviation $\sigma$ was determined from the following relation [26]:
\[
\sigma = \frac{\Delta \lambda}{2\sqrt{2\ln 2}}.
\]

Numerical integration over the spectrum was performed using a discrete set of $N_{\mathrm{spec}}=31$wavelength values uniformly distributed within the range $\lambda_{0} \pm 3\sigma$ with normalized Gaussian weights.

The spectral bandwidth range $\Delta \lambda$ extended from 0.05 to 20 nm and was discretized into 100 logarithmically spaced points. For each value of $\Delta \lambda$, a spectrally averaged hologram was generated using fixed system geometry and object parameters.

Image reconstruction from the averaged holograms was performed using back propagation at the central wavelength  $\lambda_{0}$, without taking the spectral bandwidth into account during reconstruction. This approach corresponds to typical experimental conditions and allows the effect of finite temporal coherence to be treated as an integral image degradation appearing in the reconstructed image.

The generated dataset included reconstructed images of four object types for 100 values of $\Delta \lambda$ for each object, resulting in a total of 400 images. The dataset was divided into training, validation, and test subsets with a ratio of 70/15/15.

\subsection{Neural network model}
A neural network regression model was used to estimate the source spectral bandwidth $\Delta \lambda$. The input data consisted of reconstructed images, while the output corresponded to the predicted value of $\Delta \lambda$.

Several convolutional neural network architectures were compared in this study, including baseline CNN models and their modifications with residual connections and attention mechanisms.

All models were trained using the same training protocol with fixed optimization parameters. The dataset was divided into training, validation, and test subsets as described in Section 2.3.

Model performance was evaluated on the test set using mean absolute error (MAE), mean squared error (MSE), and the coefficient of determination $\mathrm{R}^2$.

\section{Results}
\subsection{Effect of spectral bandwidth $\Delta \lambda$ on reconstructed images}

Figure 4 shows examples of reconstructed images for different values of $\Delta \lambda$, together with the corresponding dependence of the normalized RMS metric on $\Delta \lambda$. As the source spectral bandwidth increases, the image contrast decreases, and the spatial structure of the reconstructed image gradually degrades.
\begin{figure}[h]
\centering
\begin{subfigure}{\linewidth}
    \centering
    \includegraphics[width=0.2\linewidth]{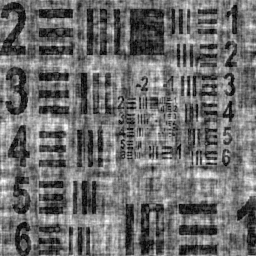}
    \includegraphics[width=0.2\linewidth]{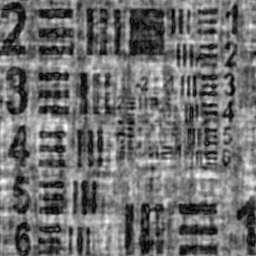}
    \includegraphics[width=0.2\linewidth]{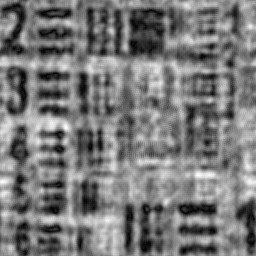}
    \includegraphics[width=0.2\linewidth]{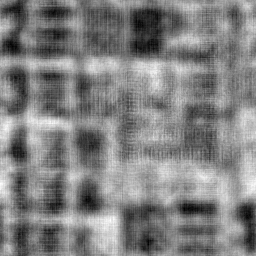}
    \caption{}
\end{subfigure}
\begin{subfigure}{0.5\linewidth}
    \centering
    \includegraphics[width=\linewidth]{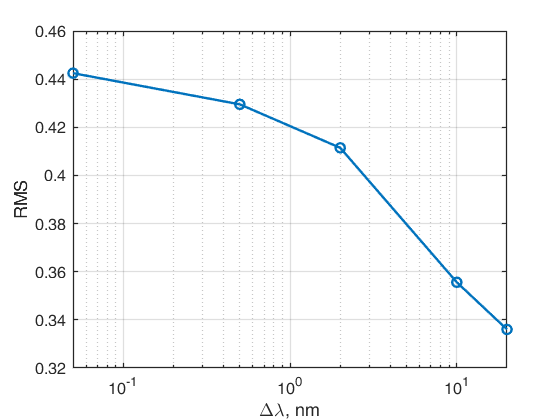}
    \caption{}
\end{subfigure}
\caption{Effect of source spectral bandwidth on reconstructed images: (a) reconstructed images of the test object for different values of $\Delta \lambda$ ($0.5, 2, 10,$ and $20$ nm); (b) corresponding dependence of the normalized RMS metric on $\Delta \lambda$.}
\end{figure}

At small values of $\Delta \lambda$, the degradation is weak and mainly appears as a reduction in local image contrast. As $\Delta \lambda$ increases, the boundaries of image features become blurred and fine details become indistinguishable. The dependence is nonlinear: only minor changes are observed at small $\Delta \lambda$, while beyond a certain threshold the degradation increases rapidly, limiting the information content of the reconstructed images.

\subsection{Neural network training results}

The study included 9 CNN-based architectures and 5 ResNet-based architectures. All models were trained using the same protocol with the Adam optimizer (learning rate $\mathrm{LR}=1\times10^{-3}$, batch size $=16$) and the Huber loss function (Smooth L1 implementation). Training was performed on the logarithmic scale $\log_{10}\left(\Delta \lambda \right)$ with Z-score normalization of the input images. Early stopping was applied based on the minimum validation MAE value (patience $=8$).

The architectures were quantitatively evaluated using mean absolute error (MAE, nm) and the coefficient of determination $\mathrm{R}^2$ calculated on the validation set. Since the $\Delta\lambda$ range extends from 0.05 to 20 nm, MAE values below 1 nm can be considered a high regression accuracy over most of the studied range.

The CBAMCNN architecture demonstrated the best results (MAE = 0.682 nm, $\mathrm{R}^2$ = 0.835). This indicates the effectiveness of attention mechanisms for identifying informative regions in degraded images. High accuracy was also achieved by the ResCNN model (MAE = 0.777 nm, $\mathrm{R}^2$ = 0.808), demonstrating the effectiveness of residual connections for this task.

Comparable results were obtained for the following architectures: ResNeXt (MAE = 0.922 nm, $\mathrm{R}^2$ = 0.785); OptimizedCNN (MAE = 0.976 nm, $\mathrm{R}^2$ = 0.770); SimpleCNN (MAE = 0.990 nm, $\mathrm{R}^2$ = 0.746); ResNet18 (MAE = 0.954 nm, $\mathrm{R}^2$ = 0.741).

It is important to note that SimpleCNN demonstrates accuracy comparable to that of more complex architectures. This indicates that the image degradation features are sufficiently pronounced for estimating $\Delta\lambda$ without significant increase in model complexity. At the same time, increasing the network depth does not lead to a substantial improvement in accuracy.

The ResNet34, ResNetD, FPNCNN, WideResNet, and SECNN architectures demonstrated MAE values in the range of 1.0–1.2 nm with $\mathrm{R}^2$ values of approximately 0.70–0.76. The SECNN model (MAE = 1.161 nm, $\mathrm{R}^2$ = 0.758) showed a relatively high $\mathrm{R}^2$ despite a larger MAE, indicating preservation of the overall dependence with increased absolute error.

The least effective architectures were MultiScaleCNN (MAE = 2.094 nm, $\mathrm{R}^2$ = 0.346) and AsymmetricCNN (MAE = 1.861 nm, $\mathrm{R}^2$ = 0.354). This indicates that the use of multiscale blocks or directionally asymmetric convolutions did not improve the model quality for the considered task. The relatively weak performance of ShallowCNN (MAE = 1.416 nm, $\mathrm{R}^2$ = 0.592) further confirms that insufficient network depth limits the stable extraction of degradation-related features.

Analysis of the validation error dynamics shows that deeper architectures often reach their minimum error at early training stages (ResNet34 — epoch 9, MultiScaleCNN — epoch 13, FPNCNN — epoch 16). Further training does not lead to significant improvement, indicating a limited effect of increasing the number of training epochs. Training dynamics for the CBAMCNN and ResCNN architectures are presented in Fig. 5.
\begin{figure}[h]
\centering
\begin{subfigure}{0.3\linewidth}
    \centering
    \includegraphics[width=\linewidth]{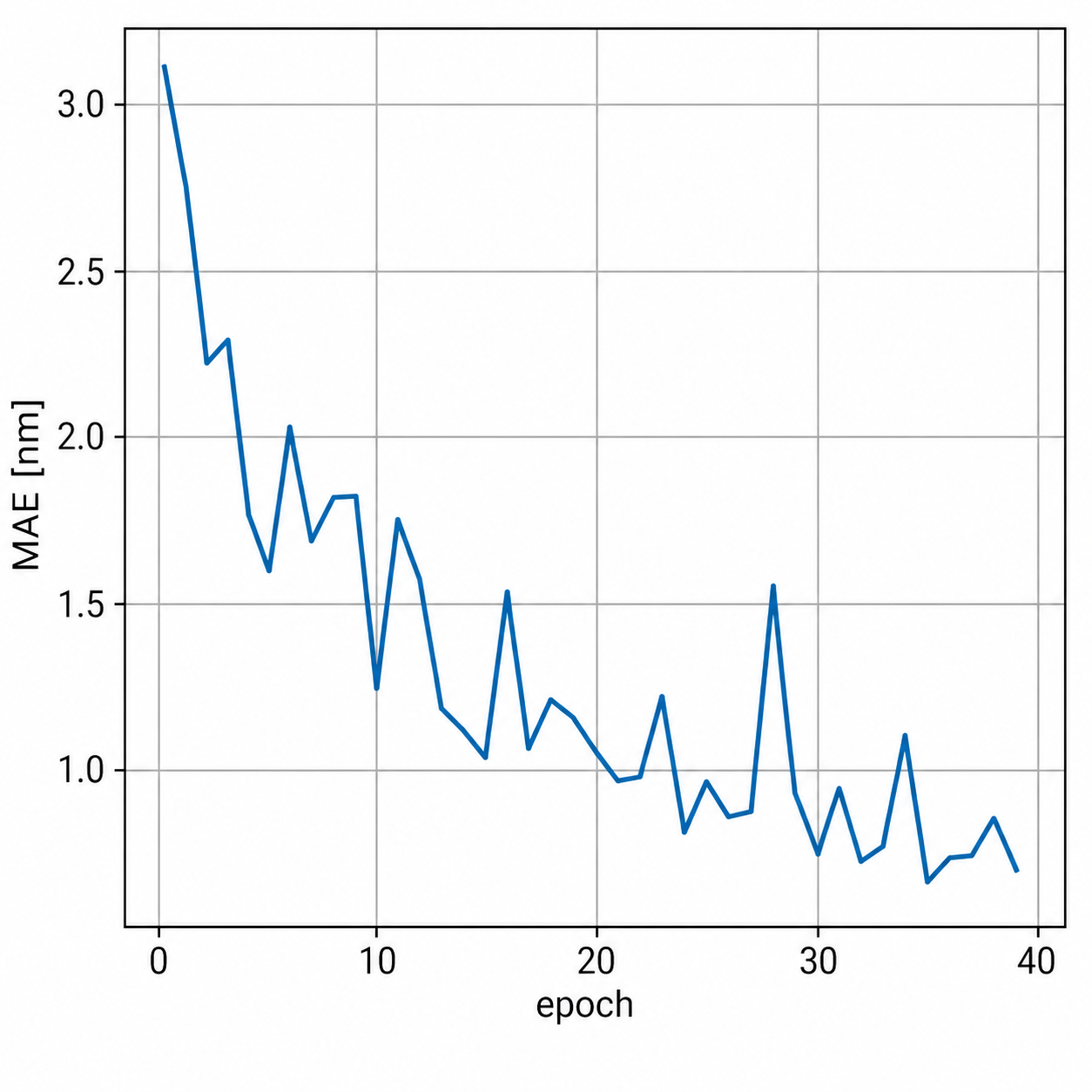}
    \caption{}
\end{subfigure}
\hspace{0.02\linewidth}
\begin{subfigure}{0.3\linewidth}
    \centering
    \includegraphics[width=\linewidth]{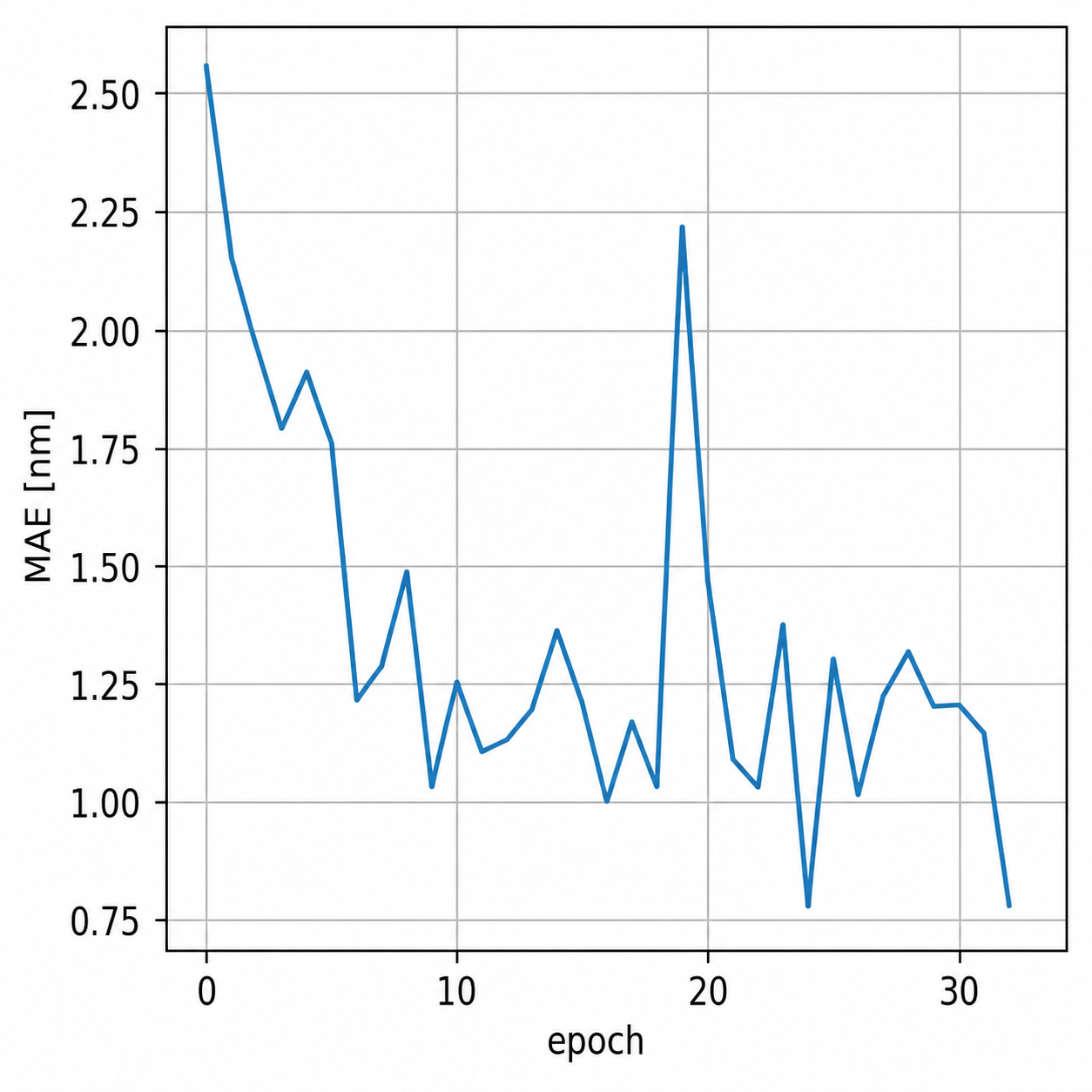}
    \caption{}
\end{subfigure}
\caption{Validation MAE dynamics during training for the architectures: (a) CBAMCNN; (b) ResCNN.}
\end{figure}

For the CBAMCNN architecture, the validation error decreases relatively smoothly with minor local fluctuations, indicating stable convergence during optimization. In contrast, the ResCNN model demonstrates more pronounced oscillations of the validation MAE, including several local spikes, which suggests lower optimization stability compared to CBAMCNN.

Thus, the CBAMCNN and ResCNN models were selected for detailed analysis as representative architectures with attention mechanisms and residual connections, respectively.

Figure 6 shows the comparison between the predicted and true $\Delta\lambda$ values, as well as the residual distributions for the CBAMCNN and ResCNN models. Both architectures demonstrate an approximately linear correspondence between the predicted and true values over the entire $\Delta\lambda$ range. In the region of small $\Delta\lambda$ values (0–3 nm), the predictions closely follow the ideal correspondence line. At larger $\Delta\lambda$ values, several outliers are observed for both models.
\begin{figure}[h]
\centering
\begin{subfigure}{0.5\linewidth}
    \centering
    \includegraphics[width=\linewidth]{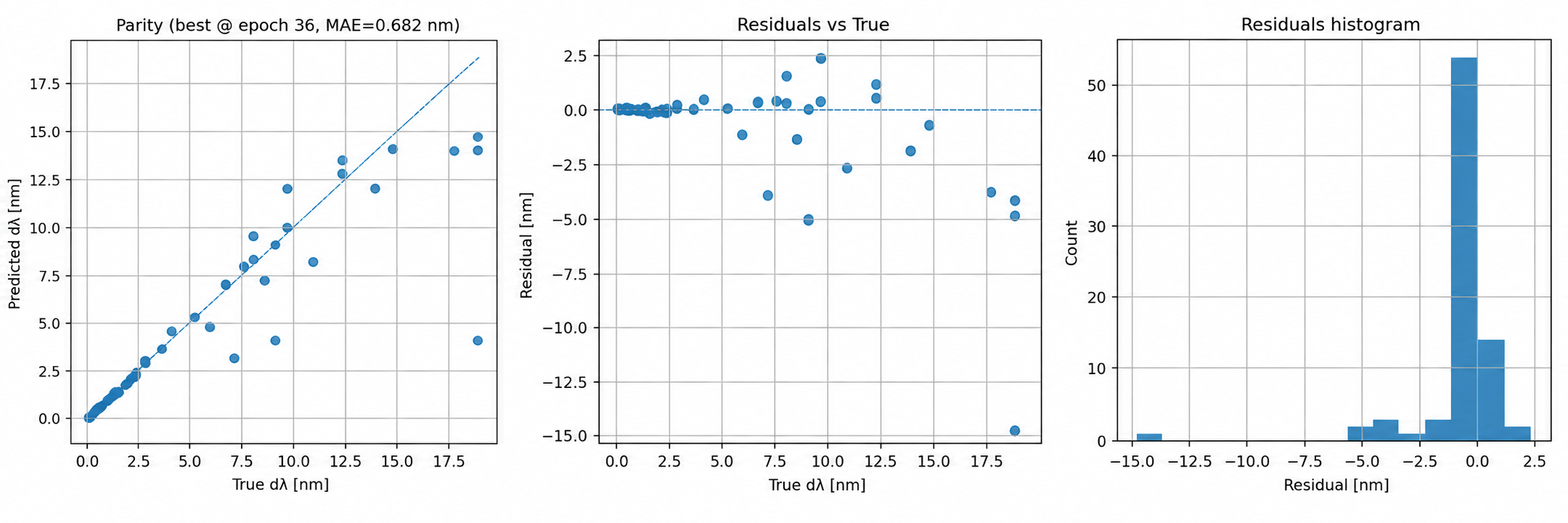}
    \caption{}
\end{subfigure}
\hspace{0.02\linewidth}
\begin{subfigure}{0.5\linewidth}
    \centering
    \includegraphics[width=\linewidth]{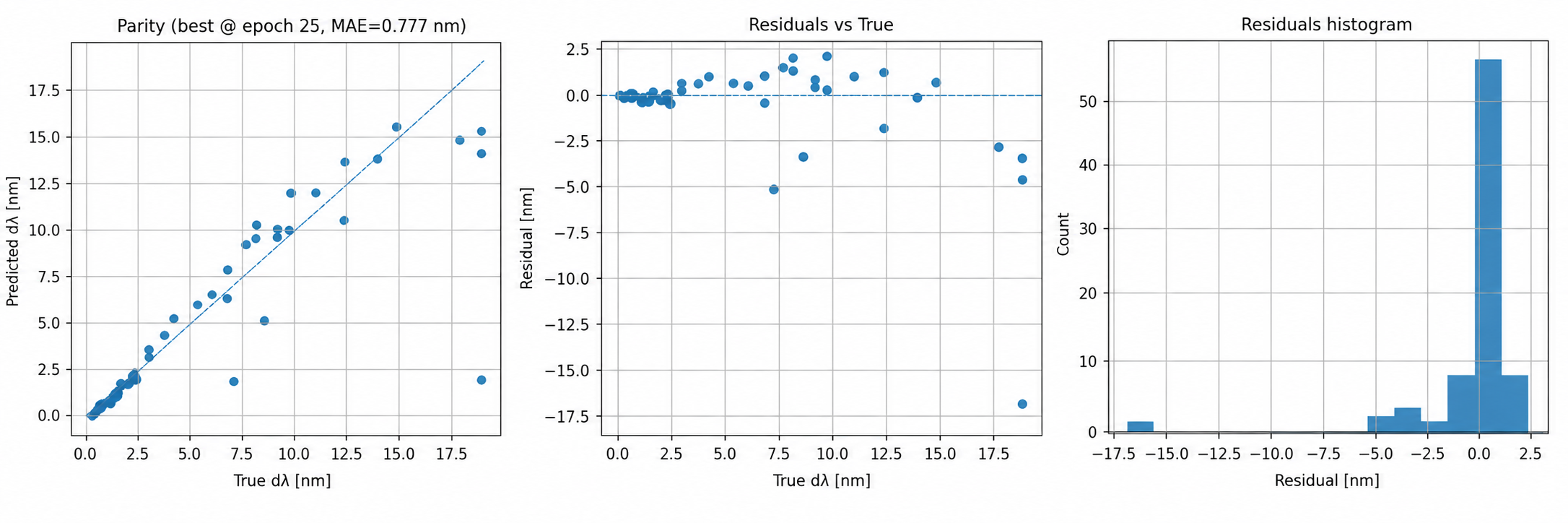}
    \caption{}
\end{subfigure}
\caption{Comparison of the predicted and true $\Delta\lambda$ values, residual distributions, and residual histograms for the models: (a) CBAMCNN; (b) ResCNN.}
\end{figure}

The residual plots exhibit a similar error distribution pattern: for small $\Delta\lambda$, the residuals remain close to zero, whereas their spread increases with increasing spectral width. No significant structural differences in the residual distributions were observed; however, the CBAMCNN model demonstrates lower overall prediction error, consistent with its smaller MAE value.

\subsection{Neural Network-Based Estimation of the Resolution of a Digital Holographic System}

To evaluate the ability of the neural network model to correctly reproduce the physical image degradation, the accuracy of the system resolution estimation was analyzed for different values of the source spectral bandwidth.

To verify the robustness of the proposed approach with respect to the choice of resolution criterion, the effective system resolution was evaluated using three independent metrics: FWHM (spatial blur width), MTF (modulation transfer function), and the USAF resolution criterion.

In the general case, the effective resolution can be considered as a quantity determined from the reconstructed image:
\[
R_{\mathrm{eff}} = R\left(I_{\mathrm{rec}}\right),
\]
where $I_{\mathrm{rec}}$ is the reconstructed image and $R\left(\right)$ is the resolution estimation operator. This formulation emphasizes that the resolution is determined not only by the system parameters, but also by the information contained in the reconstructed image.

In this work, three implementations of the operator $R\left(\right)$ were used, corresponding to different aspects of system resolution. The $R_{\mathrm{FWHM}}$ metric characterizes the spatial blur of the image and is sensitive to small resolution changes. The $R_{\mathrm{MTF}}$ metric is related to the transfer of spatial frequencies and reflects the frequency characteristics of the system. The $R_{\mathrm{USAF}}$ metric is based on the distinguishability criterion of the USAF test target and corresponds to a practical resolution assessment.

For each value of the source spectral bandwidth $\Delta\lambda$, the calibration dependencies $R_{\mathrm{FWHM}}(\Delta\lambda)$, $R_{\mathrm{MTF}}(\Delta\lambda)$, and $R_{\mathrm{USAF}}(\Delta\lambda)$ were calculated. These dependencies were used to convert the spectral bandwidth value predicted by the neural network model, $\Delta\lambda_{\mathrm{pred}}$, into the corresponding resolution estimate. For each image from the test dataset, the true and predicted resolution values were determined:
\[
R_{\text{true}}\left(\Delta \lambda_{\text{true}}\right),
\quad
R_{\text{pred}}\left(\Delta \lambda_{\text{pred}}\right).
\]
The accuracy of the resolution estimation was characterized by the absolute error:
\[
\mathrm{MAE}_R = \left| R_{\mathrm{pred}} - R_{\mathrm{true}} \right|.
\]
To compare the results obtained using different resolution metrics, the relative error was additionally used:
\[
\varepsilon_{\mathrm{rel}} =
\frac{\left| R_{\mathrm{pred}} - R_{\mathrm{true}} \right|}{\left| R_{\mathrm{true}} \right|}
\]
For the quantitative evaluation of the system resolution degradation, the FWHM metric was used, characterizing the width of the spatial blur function. The FWHM determination was performed using a standard procedure based on the analysis of the intensity profile across an image edge (Fig. 7). At the first stage (Fig. 7(a)), a region of interest (ROI) containing a sharp intensity transition was selected from the reconstructed image. Next, the edge spread function (ESF) was calculated, followed by the line spread function (LSF), obtained by differentiating the ESF (Fig. 7(b)). The FWHM value was defined as the width of the LSF at half of its maximum value (Fig. 7(c)).
\begin{figure}[h]
\centering
\begin{subfigure}{0.25\linewidth}
    \centering
    \includegraphics[width=\linewidth]{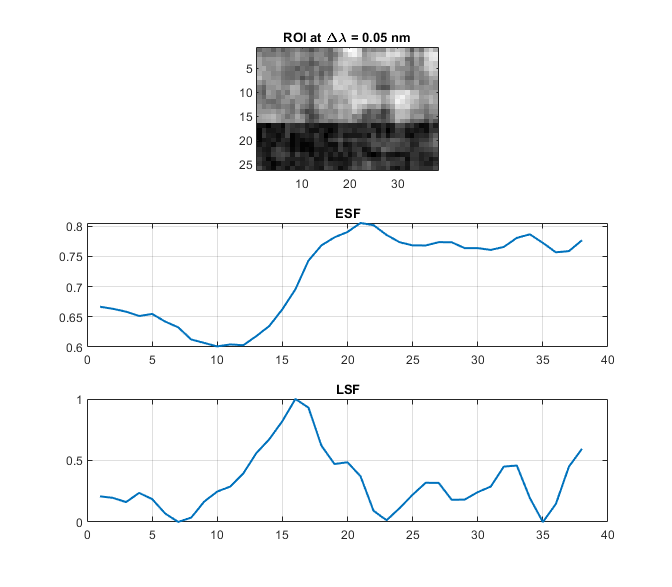}
    \caption{}
\end{subfigure}
\hspace{0.02\linewidth}
\begin{subfigure}{0.3\linewidth}
    \centering
    \includegraphics[width=\linewidth]{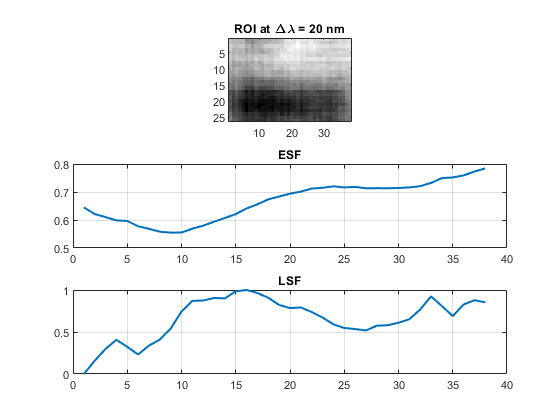}
    \caption{}
\end{subfigure}
\hspace{0.02\linewidth}
\begin{subfigure}{0.25\linewidth}
    \centering
    \includegraphics[width=\linewidth]{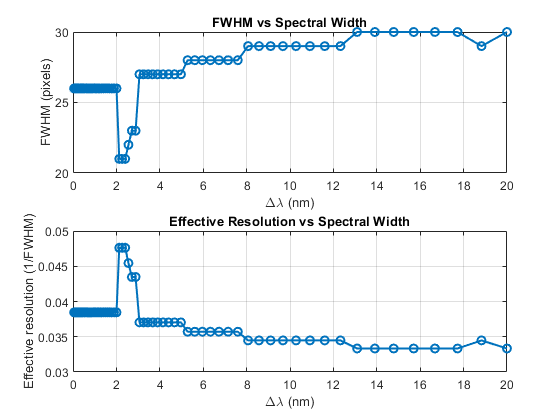}
    \caption{}
\end{subfigure}
\caption{FWHM analysis for different values of the source spectral bandwidth: (a) ROI, ESF, and LSF profiles for $\Delta\lambda$=0.05 nm; (b) ROI, ESF, and LSF profiles for $\Delta\lambda$=20 nm; (c) dependence of the FWHM and effective resolution on $\Delta\lambda$.}
\end{figure}

As can be seen from Fig. 7, increasing the source spectral bandwidth leads to a broadening of the LSF profile, corresponding to an increase in the effective spatial blur and reflecting the gradual degradation of the system spatial resolution. At small values of $\Delta\lambda$, the width of the blur function changes only slightly.

For convenience of interpretation, the effective system resolution was additionally defined as:
\[
r_{\mathrm{eff}} = \frac{1}{\mathrm{FWHM}}.
\]
As shown in Fig. 7(c), the effective system resolution decreases with increasing source spectral bandwidth. The most pronounced resolution change is observed at approximately $\Delta\lambda\approx2$ nm, corresponding to the onset of significant degradation of the reconstructed image spatial structure.

To analyze the frequency characteristics of the system, the modulation transfer function (MTF) was used, characterizing the ability of the system to transfer spatial image frequencies. The MTF was determined based on the spectral analysis of the reconstructed test-object image. In this work, the criterion MTF=0.1 was used.
\begin{figure}[h]
\centering
\begin{subfigure}{0.25\linewidth}
    \centering
    \includegraphics[width=\linewidth]{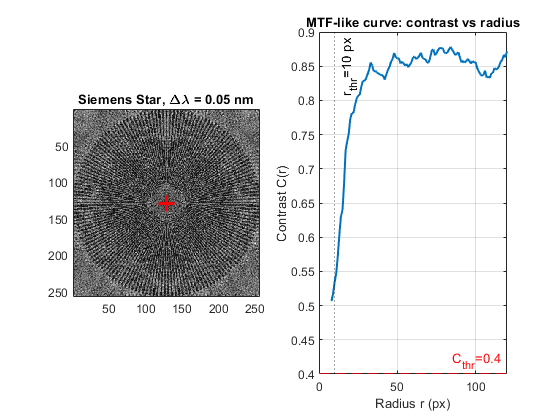}
    \caption{}
\end{subfigure}
\hspace{0.02\linewidth}
\begin{subfigure}{0.3\linewidth}
    \centering
    \includegraphics[width=\linewidth]{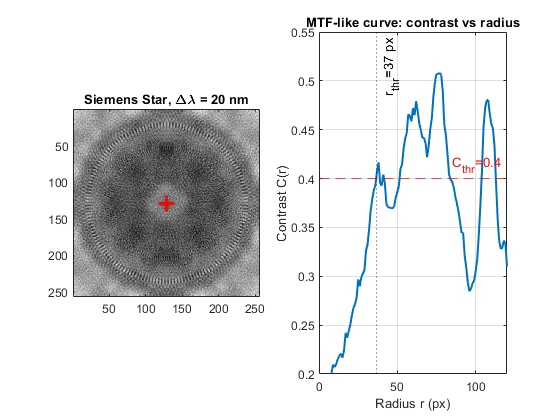}
    \caption{}
\end{subfigure}
\hspace{0.02\linewidth}
\begin{subfigure}{0.25\linewidth}
    \centering
    \includegraphics[width=\linewidth]{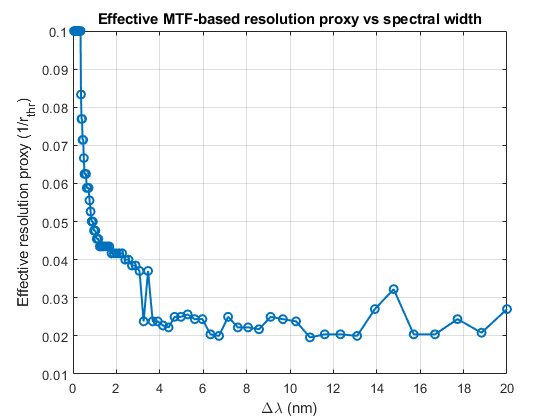}
    \caption{}
\end{subfigure}
\caption{MTF-based analysis for different values of the source spectral bandwidth: (a) contrast profile for the Siemens star test object at $\Delta\lambda$=0.05 nm; (b) contrast profile at $\Delta\lambda$=20 nm; (c) dependence of the effective MTF-based resolution proxy on $\Delta\lambda$.}
\end{figure}

Figures 8(a,b) show the MTF curves obtained for different values of the source spectral bandwidth $\Delta\lambda$. Comparison of the curves demonstrates that increasing $\Delta\lambda$ leads to a decrease in the MTF amplitude at high spatial frequencies, as well as the appearance of pronounced oscillations in the curve shape. This indicates a reduction in the contrast of fine image structures and a nonuniform transfer of high-frequency components.

Physically, this effect is caused by spectral averaging of the interference pattern at a finite source spectral width. As $\Delta\lambda$ increases, the interference fringes corresponding to different wavelengths become partially mismatched, leading to a reduction in contrast and distortion of the high-frequency image components.

For the quantitative evaluation of the system resolution, the effective spatial frequency $R_{\mathrm{MTF}}$ was used, defined as the frequency at which the MTF reaches a specified threshold level. The dependence $R_{\mathrm{MTF}}(\Delta\lambda)$ is shown in Fig. 8(c).

As shown in Fig. 8(c), the effective spatial frequency decreases with increasing $\Delta\lambda$, corresponding to degradation of the system resolution. The most pronounced decrease is observed for $\Delta\lambda$ values on the order of several nanometers, which is consistent with the results obtained from the FWHM analysis.

To evaluate the distinguishability of image elements, the USAF test target was used. The analysis was performed based on the intensity profile along selected regions of interest (ROIs) crossing the line patterns of the test target (Fig. 9(a)).

The line contrast was determined using the Michelson formula:
\[
C = \frac{I_{\max} - I_{\min}}{I_{\max} + I_{\min}},
\]
where $I_{\max}$ and $I_{\min}$ are the maximum and minimum intensity values in the image profile.
\begin{figure}[h]
\centering
\begin{subfigure}{0.3\linewidth}
    \centering
    \includegraphics[width=\linewidth]{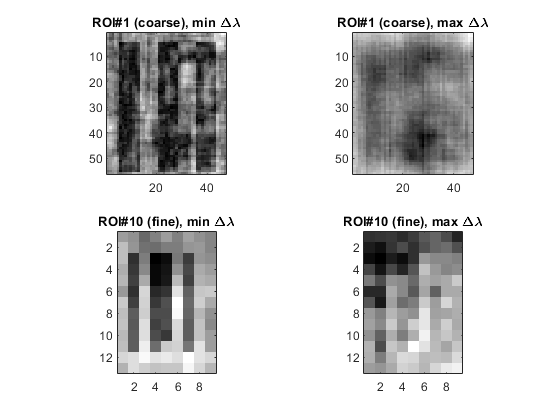}
    \caption{}
\end{subfigure}
\hspace{0.02\linewidth}
\begin{subfigure}{0.3\linewidth}
    \centering
    \includegraphics[width=\linewidth]{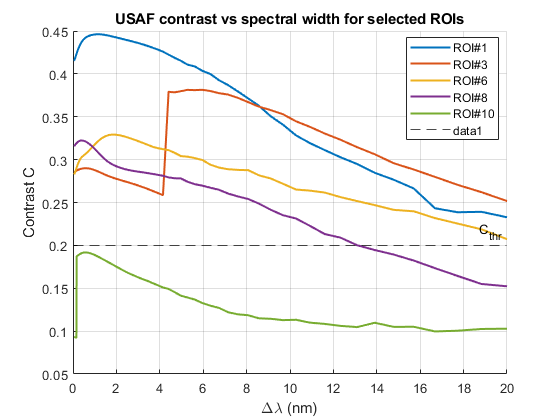}
    \caption{}
\end{subfigure}
\hspace{0.02\linewidth}
\begin{subfigure}{0.3\linewidth}
    \centering
    \includegraphics[width=\linewidth]{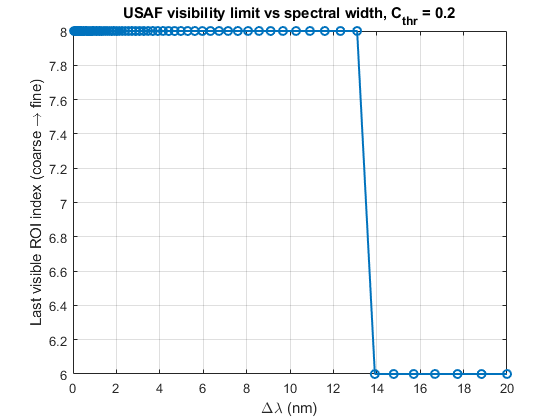}
    \caption{}
\end{subfigure}
\caption{USAF test-target distinguishability analysis for different values of the source spectral bandwidth $\Delta\lambda$: (a) examples of selected ROIs for contrast analysis; (b) contrast as a function of $\Delta\lambda$ for different ROIs; (c) limiting distinguishable spatial frequency as a function of $\Delta\lambda$.}
\end{figure}

Figure 9(b) shows the dependence of the contrast on the source spectral bandwidth $\Delta\lambda$ for different ROIs. As $\Delta\lambda$ increases, a gradual reduction in line contrast is observed. At large $\Delta\lambda$ values, the contrast becomes insufficient for reliable distinguishability of the test-target elements.

Additionally, Fig. 9(c) shows the dependence of the limiting distinguishable spatial frequency on $\Delta\lambda$. The sharp decrease of this parameter with increasing $\Delta\lambda$ corresponds to the loss of image-element distinguishability and a reduction in the effective system resolution.

Using the calibration dependencies R($\Delta\lambda$), the predicted values of the source spectral bandwidth were converted into the corresponding estimates of the system resolution. For each value of $\Delta\lambda$, the relative error of the resolution estimation was calculated.

The dependence of the relative error on $\Delta\lambda$ is shown in Fig. 10 for the three resolution metrics: FWHM, MTF, and USAF. The plots also include a 5$\%$ error threshold used to determine the critical value $\Delta\lambda_{\mathrm{crit}}$, corresponding to the applicability limit of the proposed method.
\begin{figure}[h]
\centering
\begin{subfigure}{0.3\linewidth}
    \centering
    \includegraphics[width=\linewidth]{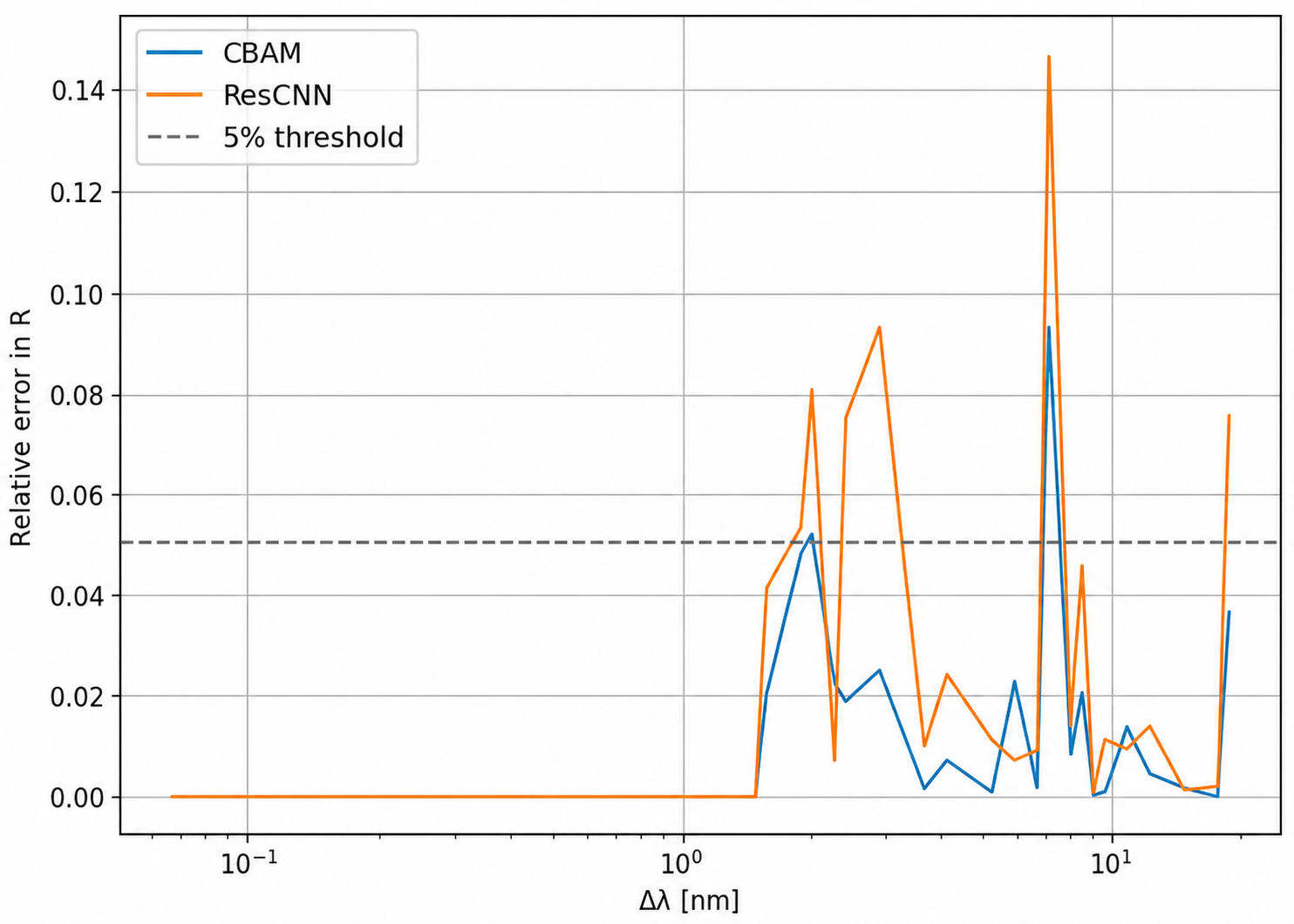}
    \caption{}
\end{subfigure}
\hspace{0.02\linewidth}
\begin{subfigure}{0.3\linewidth}
    \centering
    \includegraphics[width=\linewidth]{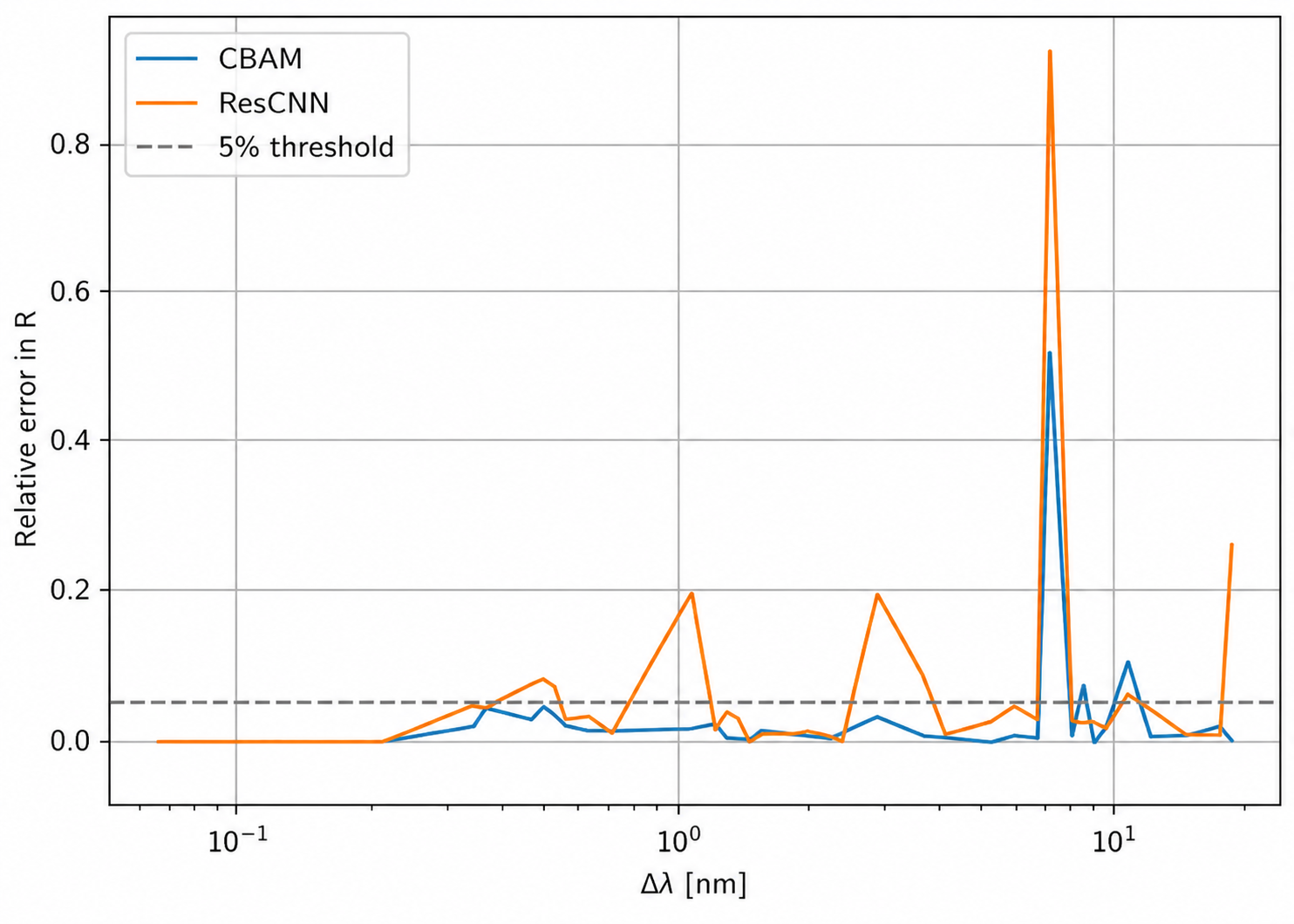}
    \caption{}
\end{subfigure}
\hspace{0.02\linewidth}
\begin{subfigure}{0.3\linewidth}
    \centering
    \includegraphics[width=\linewidth]{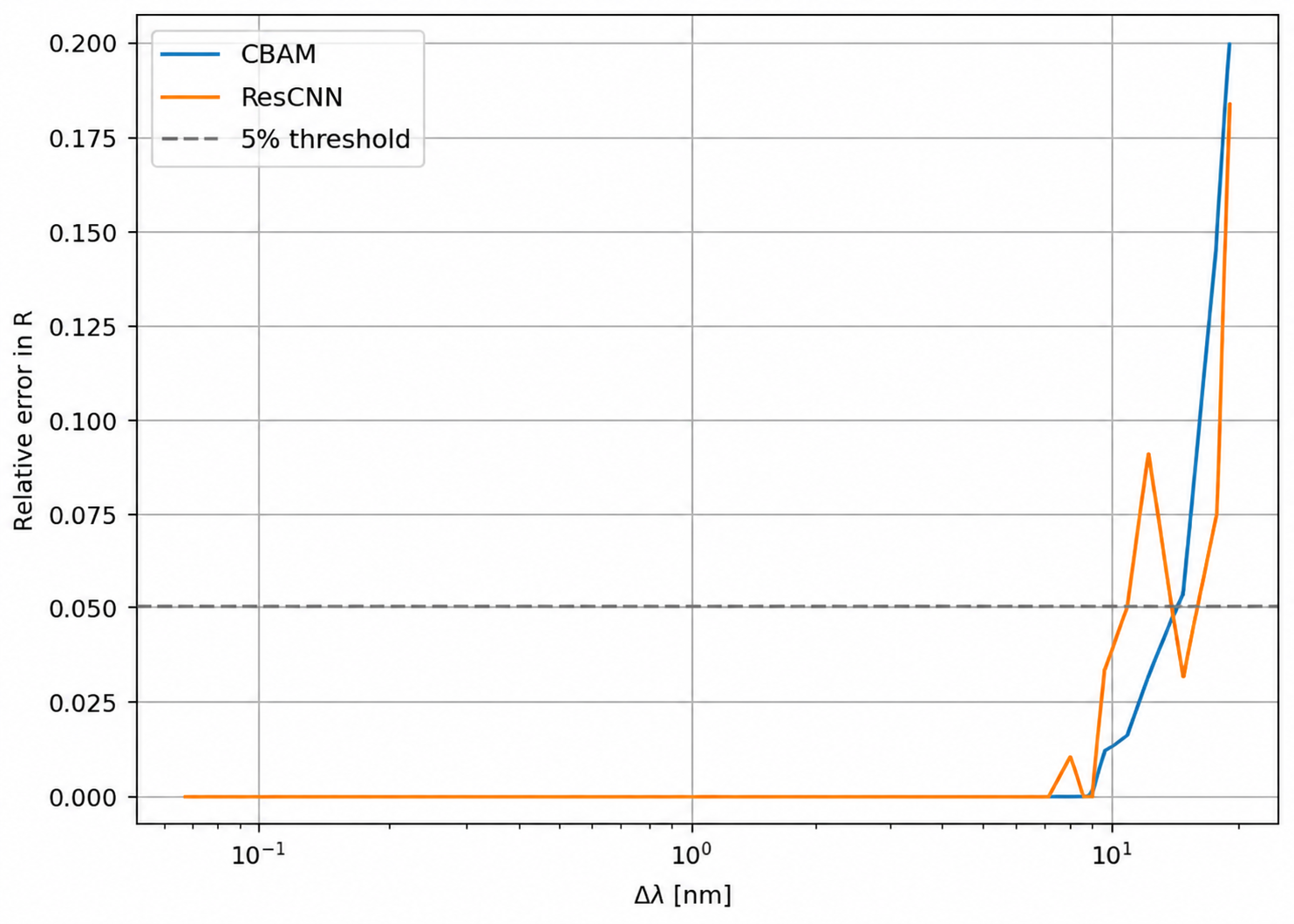}
    \caption{}
\end{subfigure}
\caption{Relative error of the resolution estimation as a function of the source spectral bandwidth $\Delta\lambda$: (a) FWHM-based error; (b) MTF-based error; (c) USAF-based error. Solid lines correspond to the CBAMCNN and ResCNN models, and the dashed line represents the 5$\%$ relative-error threshold.}
\end{figure}

For the FWHM metric, the following critical values were obtained:
\[
\Delta \lambda_{\mathrm{crit}}^{(\mathrm{FWHM},\,\mathrm{CBAM})}
\approx 2.01,
\qquad
\Delta \lambda_{\mathrm{crit}}^{(\mathrm{FWHM},\,\mathrm{ResCNN})}
\approx 1.89.
\]
The obtained values are close, indicating a weak dependence of the result on the neural network architecture.

For the MTF metric, the critical values were:
\[
\Delta \lambda_{\mathrm{crit}}^{(\mathrm{MTF},\,\mathrm{CBAM})}
\approx 7.15,
\qquad
\Delta \lambda_{\mathrm{crit}}^{(\mathrm{MTF},\,\mathrm{ResCNN})}
\approx 0.47.
\]
The large discrepancy is caused by the high sensitivity of the MTF-based metric to local errors in the estimation of $\Delta\lambda$. For the ResCNN model, an isolated outlier appears at small $\Delta\lambda$ values that is not associated with physical image degradation. Therefore, the interpretation of the MTF-based results was based on the region of stable error growth, observed for both models at approximately $\Delta\lambda\approx$6-8 nm.

For the USAF test-target distinguishability criterion, the following values were obtained:
\[
\Delta \lambda_{\mathrm{crit}}^{(\mathrm{USAF},\,\mathrm{CBAM})}
\approx 14.78.,
\qquad
\Delta \lambda_{\mathrm{crit}}^{(\mathrm{USAF},\,\mathrm{ResCNN})}
\approx 12.32.
\]
The obtained results demonstrate consistency between the neural-network-based estimation and the physical criteria of resolution degradation. The sequence
\[
\Delta \lambda_{\mathrm{crit}}^{(\mathrm{FWHM})}
<
\Delta \lambda_{\mathrm{crit}}^{(\mathrm{MTF})}
<
\Delta \lambda_{\mathrm{crit}}^{(\mathrm{USAF})}
\]
corresponds to the physical logic of image degradation: first, spatial blur increases; then, the transfer of high spatial frequencies decreases; and only at larger values of $\Delta\lambda$ does the distinguishability of image elements become lost.

The obtained results show that the proposed approach enables the estimation of the resolution of a digital holographic system under different levels of image degradation. The consistency of the results obtained using the FWHM, MTF, and USAF distinguishability criteria confirms the physical validity of the method.

\subsection{Analysis of the Model Response to Previously Unseen Types of Degradation}

Validation of the proposed approach was performed using an independent dataset containing reconstructed holographic images with different types of degradation not directly related to the source spectral bandwidth. The considered degradation factors included speckle noise, shot noise, read noise, dark current, as well as their combinations.

Unlike the training dataset, where the degradation was formed exclusively by variations in $\Delta\lambda$, the present analysis considered distortions of a different physical nature. This makes it possible to evaluate the generalization ability of the model and its sensitivity to different degradation mechanisms.

The results of the equivalent spectral bandwidth estimation, $\Delta\lambda_{\mathrm{equiv}}$, showed that the prediction accuracy strongly depends on the noise type. The smallest error was observed for images with speckle noise (mean $\approx$0.09 nm), whereas for additive noise types (shot noise, read noise, and dark current) the error increased to 0.3–1.3 nm. A relatively high error was also observed for noise-free images ($\approx$1.28 nm), indicating a systematic prediction bias in the absence of degradation.

This indicates that the model forms a representation associated with the specific physical degradation mechanism used during training. The best correspondence is observed for speckle noise, which can be explained by its interferometric nature and its similarity to the effects of spectral averaging. In contrast, additive noise does not modify the interference structure of the image; therefore, the model interprets such distortions as an equivalent increase in $\Delta\lambda$, leading to increased prediction error.

Thus, the model demonstrates selective sensitivity to the type of image degradation and extracts features associated with the interference structure rather than simply approximating the relationship between input and output. This confirms the possibility of using the neural-network-based approach for indirect estimation of optical-system parameters, provided that the degradation type corresponds to the physical model used during training.

\section{Conclusions}
The results obtained in this work demonstrate that the proposed approach for estimating the resolution of a digital holographic system based on the analysis of reconstructed images is physically justified and robust over a wide range of parameters.

Numerical simulations confirmed that an increase in the source spectral bandwidth $\Delta\lambda$ leads to nonlinear degradation of the reconstructed images, manifested as contrast reduction and deterioration of spatial resolution. The observed behavior is consistent with the physical understanding of the influence of temporal coherence on the interference structure of the field, confirming the validity of the employed model.

The neural network training results showed that the reconstructed images contain sufficient information for accurate estimation of the parameter $\Delta\lambda$. The best performance was demonstrated by architectures with attention mechanisms and residual connections, indicating the importance of efficient extraction of image-degradation features.

A key result of this work is the consistency between the neural-network-based estimation of $\Delta\lambda$ and independent physical criteria of system resolution. It was shown that the critical values of $\Delta\lambda$ determined using the neural network model are consistent with the estimates obtained from the FWHM, MTF, and USAF test-target distinguishability criteria. Moreover, the observed sequence
\[
\Delta \lambda_{\mathrm{crit}}(\mathrm{FWHM})
<
\Delta \lambda_{\mathrm{crit}}(\mathrm{MTF})
<
\Delta \lambda_{\mathrm{crit}}(\mathrm{USAF})
\]
corresponds to the physical logic of image degradation: first, spatial blur increases; then, the transfer of high spatial frequencies deteriorates; and only at larger values of $\Delta\lambda$ does the distinguishability of image structures become lost.

The presented results make it possible to interpret the neural network model not as an abstract regressor, but as a tool that extracts physically meaningful features of image degradation. This conclusion is supported by the analysis of the model generalization ability: accurate estimation of $\Delta\lambda$ is observed for distortions of an interferometric nature (for example, speckle noise), whereas additive noise does not lead to correct interpretation. This indicates the selective sensitivity of the model and its connection to the physical degradation mechanism used during training.

Thus, the proposed approach enables the transition from the analysis of individual optical-system parameters to an integral estimation of system resolution based on the reconstructed image. This is particularly important for compact and low-cost digital holographic microscopes, in which the contributions of different degradation factors are inseparable.

The limitations of the present work include the use of a synthetic dataset and the consideration of only one type of physical degradation, namely the finite temporal coherence of the source. A promising direction for future research is the extension of the model to the case of simultaneous action of multiple degradation factors, as well as validation of the proposed approach using experimental data.

\end{document}